\documentclass[a4wide]{article}

\usepackage{amsmath}
\usepackage{amssymb}
\usepackage{graphicx}
\usepackage{psfrag}
\usepackage{bm}
\usepackage{sidecap}

\begin{document}

\title{Remarks on the notion of quantum integrability}

\author{Jean-S\'ebastien Caux and Jorn Mossel \\
%
Institute for Theoretical Physics, Universiteit van Amsterdam, \\
Science Park 904, Postbus 94485, \\
1090 GL Amsterdam, The Netherlands
}

\maketitle

\begin{abstract}
We discuss the notion of integrability in quantum mechanics. 
Starting from a review of some definitions commonly used in the literature,
we propose a different set of criteria, leading to a classification of models in terms of different integrability classes.
We end by highlighting some of the expected physical properties associated to
models fulfilling the proposed criteria.
\end{abstract}

\newpage

\section{Introduction}
\label{sec:Introduction}

Classical mechanics is a subject with a unique level of maturity:  it is one of the most enjoyable
to learn, and through the beauty and powerfulness of its formalism is often considered as a prototypical 
example of `how things should ideally be done' in physics. One of the most powerful concepts in the
study of the dynamics of classical systems is the notion of integrability (in the sense of Liouville, see
{\it e.g.} \cite{ArnoldBOOK}), namely that if a system with $n$ degrees of freedom ({\it i.e.} with
$2n$-dimensional phase space) possesses $n$ independent first integrals of motion in involution ({\it i.e.} Poisson-commuting), then
the system is integrable by quadratures. The meaning of `integrable' here is thus transparent,
namely that the differential equations describing the time evolution can be explicitly integrated 
using action-angle variables. The solutions of the equations of motion thus display periodic motion
on tori in phase space, and ergodicity is absent, in contrast to non-integrable models which explore
phase space densely in the course of time. Besides providing explicit solutions to the time evolution,
the classical notion of integrability thus partitions classical models into separate classes of
integrable and non-integrable models with manifestly different physical behaviour.

It thus comes as a surprising (and insufficiently known) fact that translating the notion of integrability 
to the quantum context has faced numerous pitfalls, and remains to this day a subject of debate. 
This leads to some unfortunate widespread confusion, since integrability is mentioned very often in contemporary 
discussions and publications concerning among other themes in- and out-of-equilibrium dynamics, relaxation and thermalization 
of many-body quantum systems under current theoretical and experimental investigation. If quantum
integrability is ill-defined, how can we thus invoke it at all?

Questioning the precise meaning of `quantum integrability' has been done on many occasions.
Nearly two decades ago, in an eminently readable article, 
Weigert \cite{1992_Weigert_PD_56} summarized some fundamental issues and discussed the shortcomings 
of commonly used definitions. Delving further into the details is however not usually done
in research articles, but rather in private discussions or proceedings of lectures given by
eminent researchers in the field, {\it e.g.} \cite{2004_Faddeev_HenriPoincare}. 
It is also a subject of ongoing work (see for a recent example \cite{2009_Clemente_Gallardo_IJGMMP_6}).
Since quantum integrability was very often mentioned and discussed by many participants 
during the StatPhys 24 conference, we found it appropriate to use the occasion offered by these proceedings
to share a few hopefully worthwhile thoughts, observations, suggestions and conclusions on this important theme.

The paper is organized as follows. We first put the problem in context, highlighting precisely {\it what}
the problem is, and what kind of solution would ideally be required. In Section \ref{sec:OldDefinitions},
we review many definitions commonly found in the literature, and 
collect our thoughts and comments on each of them. After summarizing the conclusions reached, we propose
a new categorization in Section \ref{sec:Definition}, and provide examples of where known models fit within our scheme
in Section \ref{sec:List}. 
The physical consequences of our definition are discussed in Section \ref{sec:Physics}, which is followed
by our conclusion.

\section{Motivations}
\label{sec:Motivations}
We can begin with the simplest amd most important questions: Don't we have a proper
definition already? Why is this question interesting and important? 
The first question will be answered in the negative in the following section. The second question
is best answered by reflecting on the classical case: since the presence or absence of integrability
in a classical system is associated to such drastic differences in physical behaviour, the lack
of a proper understanding of the quantum equivalent inevitably means that we must be `missing out' on
some important properties and features; the lack of a quantum equivalent to the KAM theorem \cite{1954_Kolmogorov_DANSSSR_98,1963_Arnold_UMN_18,1962_Moser_NAWG_1} (on the stability of quasi-periodic motion in the presence of small perturbations) is possibly
the most striking illustration of this point, and makes it difficult to extract hard statements on
the equilibration and thermalization of many-body quantum systems.

The lack of correspondence between classical and quantum integrability, which we will discuss further
below, leads us to ask more basic questions about the differences between classical and quantum systems \cite{1984_Hietarinta_JMP_25}.
A first point worth remembering is that quantum mechanics differs markedly from classical mechanics 
in the way it counts degrees of freedom. In quantum mechanics, discretization of levels means that we can 
comfortably work with finite-dimensional Hilbert spaces: spins, bound atomic levels, {\it etc.} have eigenstates 
which we can label with a discrete quantum number taking a finite set of values, and we typically say that
the number of degrees of freedom of a quantum system is the dimensionality of its Hilbert space.
By contrast, in classical mechanics, we count degrees of freedom by specifying how many pairs of conjugate
phase space variables are necessary to specify the configuration of a system. Each variable can take
on a continuum of values. 
In any quantum-classical correspondence, we would thus associate the number of 
classical degrees of freedom to the multiplicity of infinities of the dimension of the Hilbert space.
There thus cannot be a classical equivalent to a quantum system with a finite-dimensional
Hilbert space, and this already means that classical integrability is insufficient as a basis for defining quantum
integrability in general. 

When thinking about conserved charges, the notion of Liouville integrability includes a specification of how many independent
charges we need, namely a number identical to the number of degrees of freedom $n$ (in which case it is said
that the system possesses a complete set of charges, or is completely integrable). If we can provide more than
$n$ charges, the system is said to be superintegrable; if we can produce $2n$ charges, the system 
is maximally superintegrable (see {\it e.g.} \cite{2002_Gravel_JMP_43} and references therein).  
One semantic pitfall is associated to the word `complete'. Namely, 
one fundamental notion in quantum mechanics is that of a complete set of commuting observables (CSCO), namely
a set of commuting quantum operators whose eigenvalues are sufficient to uniquely specify a state in Hilbert space. 
In the context of integrability, the word `complete' takes on a different meaning: for a fully nondegenerate system, a single operator (the
Hamiltonian) already forms a CSCO. The cardinality of a CSCO is thus patently not the number of conserved charges
we should be looking for in the quantum case. We should of course be looking for a maximal abelian subalgebra of
quantum operators in Hilbert space, meaning that we should be able to display a number of commuting operators coinciding
with the dimensionality of the Hilbert space in order to call our set of charges `complete'. To avoid this pitfall, we
will thus rather talk about `maximal' sets than `complete' ones.

Before going further with our discussion of quantum integrability, it is worthwhile to follow 
the example of \cite{1992_Weigert_PD_56} and formulate
a number of requirements for a meaningful and useful definition of this concept.
Most importantly, 
\begin{enumerate}
\item it should be unambiguous;
\item it should partition the set of all possible quantum models into distinct classes;
\item different classes of models should display distinguishable physical behaviour.
\end{enumerate}
In addition to these, we could formulate a number of extra requirements, namely:
a) the contact with the classical limit should be natural;  b) the contact with integrable field theory should be natural; 
c) the different classes should be (algorithmically) distinguishable, 
{\it i.e.} it should be easily feasible to determine which class a model belongs to, etc.
These are however less crucial criteria than the ones we have selected.

\section{Common definitions used in the literature}
\label{sec:OldDefinitions}

In this section, we initiate our discussion by summarizing a number of definitions of 
quantum integrability encountered in the literature. We briefly comment each one. 

\paragraph{QI:N}
{\it A system is quantum integrable {\bf QI:N} if it possesses a maximal set of independent 
commuting quantum operators ${\cal Q}_\alpha$, $\alpha = 1, ..., \mbox{dim}({\cal H})$.} \\

Allowing for a bit of flexibility in the precise terms used, this is (at least in spirit) 
overwhelmingly the most common definition of quantum integrability encountered in the literature.
It has the appeal of being directly related to the classical notion of integrability, in the sense
of being essentially a word-for-word translation after replacing Poisson brackets with commutators.

Definition {\bf QI:N} is given the label {\bf N} for a simple reason: it is too naive.
Its fatal flaw is absolutely trivial:  all quantum models associated to (limits of) finite-dimensional Hilbert spaces fall
under the label {\bf QI:N}. By the spectral theorem, all Hermitian Hamiltonians are readily diagonalizable;
one thus obtains $\mbox{dim}({\cal H})$ orthogonal state vectors $| \Psi_\alpha \rangle$ from which one can build
projectors ${\cal Q}_\alpha = | \Psi_{\alpha} \rangle \langle \Psi_\alpha|$, the set of which constitutes a
maximal independent commuting set. So is every quantum system we can think of to be called integrable? Well, a court jester
would amuse himself doing precisely this, but we have to reject this pathway by invoking one of the requirements we
had about a proper definition, namely that it should separate models into distinct classes. Definiton {\bf QI:N}
blatantly fails to do so, and is thus to be rejected as being formally useless.

Loopholes of definition {\bf QI:N} were also extensively discussed in \cite{1992_Weigert_PD_56}. Loophole (A)
there corresponds to the flaw mentioned above. A second loophole mentioned is associated to the notions of `maximal' and `independent'
quantum operators: a theorem of von Neumann \cite{1931_vonNeumann_AM_32} 
states that it is possible to 
encode any number of commuting Hermitian operators into a single Hermitian operator ${\bf Q}$ (in other words,
any operator ${\cal Q}_\alpha$ can then be viewed as a function ${\cal Q}_\alpha = f_\alpha ({\bf Q})$), so the very basic notion of
the {\it number} of independent operators actually seems ill-defined. Going further, since we allegedly can't even properly count
the number of charges we have, we are then prevented from honestly declaring that a set be maximal. 

While we cannot repair the first fatal flaw of definition {\bf QI:N}, let us address the second point in more detail. 
We do not dispute the validity of von Neumann's theorem, however we do not agree that it is of relevance here.
First of all, thinking about the counting of conserved charges which can be defined as how one would represent them
in terms of matrices in the eigenbasis, there is no doubt that the number required to call a set maximal
coincides with the dimensionality of 
Hilbert space $\mbox{dim}({\cal H})$, since this is the number of independent diagonal entries. 
As far as counting is concerned, the notion of algebraic independence 
(that is, the set of charges does not obey any nontrivial polynomial equation) 
is sufficient to make it well-defined, and is already in use in the literature (see for example 
\cite{2002_Gravel_JMP_43} and references therein).

We thus have to look for something beyond the naive definition. Of course, one of the main assumptions was
that in practice, we {\it could} actually diagonalize the Hamiltonian to obtain the charges as projectors.
This suggests a more pragmatic definition:

\paragraph{QI:ES}
{\it A system is quantum integrable {\bf QI:ES} if it is exactly solvable, in other words if we can construct
its full set of eigenstates explicitly.} \\

While this reminds us of the action-angle variables in the classical case, 
the reader will probably agree that this washed-down definition does not take us very far. 
This definition could be further categorized according to which method is employed to obtain
the eigenstates:  Fourier transform for free theories, Bethe Ansatz for specific models (although 
the completeness of the set of Bethe eigenstates is not formally proven for all models), etc.
We also reject this definition on the grounds that it does not fulfill all our criteria:  the
third, in particular, is hard to relate to.

\paragraph{QI:HO} {\it A system is quantum integrable {\bf QI:HO} if it can be mapped to
harmonic oscillators.} \\

 This is not really practical: such a mapping is hard to construct 
explicitly, even for models we know how to solve exactly. Anyway, the existence of such
a mapping is guaranteed by the fact that all $C^*$ algebras are unitarily equivalent.

A related definition is the one used in \cite{1995_Weigert_CSF_5}, where spin systems are
classified as integrable if there exists an operator mapping leading to a Hamiltonian 
expressible in terms of action operators ({\it i.e.} $\hat{J}_z$). 

In the mathematical literature, there exists a so-called algebro-geometric definition of quantum complete
integrability \cite{1997_Braverman_TG_2}, implying that a system is algebraically integrable ({\it i.e.} 
its eigenvalue problem is solvable by quadratures, mirroring the classical case).
This allows nice things such as the application of the logic of Darboux transformations in the
quantum context (see {\it e.g.} \cite{1999_Horozov_LMP_49}) to generate nontrivial models.  
This definition applies naturally to quantum models
with continuous degrees of freedom (the notion of `complete' is then like in the classical case),
but we find it to be too close to the classical definition to be sufficiently general.

Instead of looking at exact solvability, the way out might be to look more precisely at the nature of the wavefunctions. This is in
particular what the Bethe Ansatz teaches us:  the scattering in one-dimensional integrable theories has the
remarkable feature of being factorizable in two-body scatterings. This forms the basis for
Sutherland's definition:

\paragraph{QI:ND}
{\it A system is quantum integrable {\bf QI:ND} if the scattering it supports 
is nondiffractive \cite{SutherlandBOOK}.} \\

This definition is very appealing. Although it does not directly mention the existence or the number of 
conserved charges, one can easily construct those from {\it e.g.} the set of conserved momenta. 
It relates more or less directly to the classical definition, 
but is also well-defined, testable (at least in principle), and physically meaningful
in the sense that nondiffractive scattering can be translated directly into non-ergodicity.
It encompasses relativistic integrable field theory, where this nondiffraction condition is 
simply the factorization condition.

One shortcoming is that it relies on observing the effects of scattering in the `asymptotic region' in 
the space of states (where all particles are clearly separated), and is thus only suited for models defined 
directly in the continuum. While we find this definition the best one available, we would rather
find nondiffractive scattering as a {\it consequence} of integrability, rather than as its defining
feature. \\

We can go further and try to formulate other `physical' definitions, for example

\paragraph{QI:ELS} {\it A system is quantum integrable if its energy level statistics is Poissonian.} \\

Looking at the energy level statistics of quantum models is another very appealing way to
address the question of integrability, since it directly connects with classical mechanics
via Berry and Tabor's semiclassical reasoning \cite{1977_Berry_PRSLA_356}.  This
showed that the energy level statistics of generic quantum integrable systems is Poissonian.
Since this is based on semi-classical reasonings, only quantum systems with continuous degrees of freedom
(and thus infinite-dimensional Hilbert spaces) are concerned. 
This was however further investigated for quantum lattice models \cite{1993_Montambaux_PRL_70,1993_Poilblanc_EPL_22,2004_Rabson_PRB_69}
such as Heisenberg, $t-J$ and Hubbard-like models, explicitly verifying that for nonintegrable cases
the statistics becomes Wigner's GOE. 
For Richardson-Gaudin-like models, the degree of freedom afforded by the large number of tunable internal
parameters means that essentially any distribution can be mimicked; however,  these
quickly become Poissonian upon turning the interaction on \cite{2004_Relano_PRE_70}. 
We also refer the reader to the recent preprint \cite{1004.4844} for recent examples and
more extensive literature citations. A related definition is 

\paragraph{QI:LC} {\it A system is quantum integrable if it shows level crossings 
({\it i.e.} does not show level repulsion).} \\

Here, the absence of level crossings in {\it e.g.} numerical solutions of finite models is then interpreted as
an observation of non-integrability (see for example \cite{2008_Stepanov_PRE_77} and references therein). 
This definition really pertains to families of models with a tunable parameter, and again cannot be sufficiently
general. \\

Overall, we would rather view these features of energy level distributions used in
{\bf QI:ELS} and {\bf QI:LC} as consequences of 
integrability, rather than as definitions. In any case, there exist counterexamples, for example, the Richardson-like models
mentioned above, and Haldane-Shastry type models 
\cite{2005_Finkel_PRB_72,2006_BasuMallick_NPB_757,2008_Barba_EPL_83,2008_Barba_PRB_77,2009_Barba_NPB_806,2010_Barba_NPB_839,2010_Enciso_PRE_82}, 
the main conclusion there being that the energy level statistics was neither Poissonian nor of Wigner type.
There is thus no universally acceptable definition of the form {\bf QI:ELS}; on the other hand, 
{\bf QI:LC} formally only makes sense when discussing models with tunable parameters, which cannot cover all cases of interest.

\subsection{Where does this leave us?}
Let us attempt a summary of the important lessons of the preceding discussion.
Most importantly, the mere existence of conserved charges (which for any model with a finite Hilbert space is a trivial 
consequence of the spectral theorem) cannot be a sufficiently meaningful criterion to 
classify quantum models in different classes. It is therefore imperative to look into the actual 
structure of the conserved charges when trying to distinguish classes of quantum models. But the
structure of quantum operators really depends on the representation we use to write them down.
We can always choose to work in the Hamiltonian eigenbasis, in which the charges are diagonal, 
and we can even define them such that they have a single matrix entry (this is the case of the projectors 
onto eigenstates).  Therefore, the existence of a basis in which the charges look simple again does not provide
us with a classification criterion, and a practical definition of quantum integrability cannot exist 
which does not identify a `preferred' basis 
in which the structure of the charges must be looked at. 
In view of our (admittedly, optional) desire to keep a semblance of quantum-classical correspondence, 
we could restrict {\it e.g.} to `real space' and `momentum space' bases,
or more generally accept that a definition of quantum integrability is only valid in a specified basis.

As we have mentioned before, the counting of the number of conserved charges is also a subtle point leading to various pitfalls,
as our discussion about von Neumann's theorem illustrated. A further pitfall occurs 
when counting is mentioned together with a physically meaningful concept, that of locality:
there is a simple contradiction in terms when talking about a `maximal' (or the unfortunately more commonly used `complete')
set of `local' conserved charges. Enforcing `locality' restricts the number of operators which can be defined to a number
which is negligible in comparison to the required number for `maximalness' (the dimensionality of the Hilbert space).
It is thus patently impossible to have a maximal set of local charges, and one of the two concepts has to give way:
either our required set is not maximal, or it contains nonlocal charges.  
This confusion probably originates from careless interpretations of the integrable quantum field theory literature, 
by confusing an infinite set of charges with a maximal one. This is also mentioned in the
context of (the infinite-size limit of) integrable lattice models: for example, in the fundamental paper \cite{1979_Babbitt_JMAA_72},
the set of conserved charges for the infinite Heisenberg chain was shown to form a maximal abelian family, and this
was equated (first paragraph of main text) to an `{\it explicit, complete} set of mutually commuting operator invariants which have
local densities...'. We take issue with this use of words, and prefer to rephrase this in terms of the weaker statement that
there then exists an infinite number of conserved charges which can be written as integrals of local densities, without
specifying that this set is `maximal' (it is not; it only is `maximal' with respect to local operators). 
This leaves us with the question: how many charges do we need? The lesson from the earlier discussion is that
we can probably drop the `maximality' requirement and still make meaningful statements.

With this in mind, let us now attempt to formulate a definition which takes care of all these issues.

\section{Alternate definition of quantum integrability}
\label{sec:Definition}
We define a {\it size sequence} as an infinite sequence of strictly increasing integers $(N_1, N_2, N_3, ...)$, $N_1 < N_2 < N_3 < ...$.
To each given size $N_a$ in the size sequence, we associate
a Hilbert space ${\cal H}^{(N_a)}$ obtained from tensoring $N_a$ elemental Hilbert spaces
${\cal H}_j$, $j = 1, ..., N_a$.  We assume that each elemental Hilbert space is finite-dimensional, 
$\mbox{dim}({\cal H}_j) \equiv d_j < \infty$, so $\mbox{dim}({\cal H}^{(N_a)}) = \prod_{j=1}^{N_a} d_j \equiv d^{(N_a)}$ is also finite.
As we move up the size sequence, we assume that we simply tensor in additional elemental spaces,
${\cal H}^{(N_{a+1})} = {\cal H}^{(N_a)} \otimes_{j = N_a + 1}^{N_{a+1}} {\cal H}_{j}$.\footnote{Contrary to what might
be assumed, there is no implicit restriction to one-dimensional systems.}

Operators in ${\cal H}_j$ can be represented by $d_j \times d_j$ Hermitian matrices, which can be decomposed in a basis
${\bf e}_j^i$, $i = 1, ..., d_j^2$.
Operators ${\cal Q}^{(N_a)}$ in ${\cal H}^{(N_a)}$ can thus be decomposed in the 
$(d^{(N_a)})^2$ basis matrices ${\bf e}^{i_1 ... i_{N_a}} \equiv \otimes_{j=1}^{N_a} {\bf e}_j^{i_j}$ 
(which we call the {\it preferred basis}), 
${\cal Q}^{(N_a)} = \sum_{i_1, ..., i_N} {\cal Q}^{(N_a)}_{i_1 ... i_{N_a}} {\bf e}^{i_1 ... i_{N_a}}$.
For a given operator ${\cal Q}$ in a given preferred basis ${\bf e}^{i_1 ... i_{N_a}}$, we denote the number of nonzero
entries ${\cal Q}^{(N_a)}_{i_1 ... i_{N_a}}$ as ${\mathbb N}_{\bf e} ({\cal Q}^{(N_a)})$.
For a size sequence of operators $({\cal Q}^{(N_1)}, {\cal Q}^{(N_2)}, {\cal Q}^{(N_3)}, ...)$, 
we define the {\it density character} as the nature\footnote{In our definition, the terminology we use for $O(f(N))$ is identical to that used for classifying algorithmic time complexity. We can thus talk about {\it linear}, {\it sublinear}, {\it quasi-polynomial}, {\it subexponential}, {\it etc}. If needed, a term proportional to the unit operator should be subtracted when establishing the density character of an operator.} of the minimal function $f(N_a)$ such that
${\mathbb N}_{\bf e} ({\cal Q}^{(N_a)}) < f(N_a)$ $\forall ~a$.

We can now proceed to define a corresponding size sequence of Hermitian operators $(H^{(N_1)}, H^{(N_2)}, H^{(N_3)}, ...)$, which
we interpret as Hamiltonians acting in their respective Hilbert space ${\cal H}^{(N_a)}$.  
We assume that the Hamiltonians in the size sequence are built according to some simple and meaningful algorithm
${\cal A}: {\cal H}^{(N_a)} \rightarrow H^{(N_a)}$.
We restrict to 
Hamiltonians having polynomial density character, in other words to Hamiltonians which can be
expressed in terms of a finite number of sums over elemental labels.

Being Hermitian and finite, each Hamiltonian $H^{(N_a)}$ obeys the spectral theorem and therefore possesses 
a complete set of eigenvectors. This also means that it {\it automatically} possesses a 
maximal set of conserved charges $\{ {\cal Q}^{(N_a)}_{\alpha} \}$, 
$\alpha = 1, ..., d^{(N_a)}$ in involution.  We can arrange the labelling of the charges in such a way that
meaningful finite size sequences $({\cal Q}^{(N_a)}_\alpha, {\cal Q}^{(N_{a+1})}_\alpha, {\cal Q}^{(N_{a+2})}_\alpha, ...)$, 
for all individual ${\cal Q}^{(N_a)}_\alpha$ with
$\alpha = 1, ..., \alpha_{max} \leq d^{(N_a)}$ can be defined.

\paragraph{Definition:} we will call a Hamiltonian $H$ $O(f(N))$ {\it quantum integrable}
if it is a member of a sequence
$(H^{(N_1)}, H^{(N_2)}, H^{(N_3)}, ...)$ of operators having $O(f(N))$ density character in the preferred basis, for which
it is possible to define a sequence of sets of 
operators $(\{ {\cal Q}^{(N_1)} \}, \{ {\cal Q}^{(N_2)} \},
\{ {\cal Q}^{(N_3)} \}, ...)$ such that
\begin{enumerate}
\item all operators ${\cal Q}^{(N_a)}_\alpha$ in $\{ {\cal Q}^{(N_a)} \}$ commute with each other and with their Hamiltonian $H^{(N_a)}$;
\item the operators in $\{ {\cal Q}^{(N_a)} \}$ are algebraically independent;
\item the cardinality ${\cal C}^{(N_a)}$ of the set $\{ {\cal Q}^{(N_a)} \}$ becomes unbounded  
in the infinite size limit\footnote{Notice the fact that we do not require the set of conserved charges to be maximal.};
\item each member ${\cal Q}^{(N_a)}_\alpha$, $\alpha = 1, ..., {\cal C}^{(N_a)}$ 
of the set $\{ {\cal Q}^{(N_a)} \}$ can be embedded within a 
sequence of operators $({\cal Q}^{(N_1)}_\alpha, {\cal Q}^{(N_2)}_\alpha, {\cal Q}^{(N_3)}_\alpha, ...)$
with $O(f(N))$ density character in the preferred basis.
\end{enumerate}

We can thus talk about models in the linear, polynomial, quasi-polynomial, etc. integrability classes.
For the preferred basis, while our formulation is purposefully left generic, we are of course mainly 
thinking about a real-space based basis (in this case, the Fourier basis is also automatically compatible).

The reader might feel that this definition falls a bit `out of the blue'. We will attempt to
make its meaning clearer by first specifying which classes known models fall into, and thereafter
discussing in more physical terms how this definition should be understood.

\section{Which integrability class do known models belong to?}
\label{sec:List}
Having suggested to classify quantum models into different categories, our task now turns to 
giving explicit examples.

\subsubsection*{Free theories}
There is little doubt that free theories should be considered integrable in some way, and that they should be the
simplest types of integrable models available.
Imagine thus that we have a lattice model for a set of particles, with specified hopping amplitudes 
(with no restriction on dimensionality or on these being
local and/or long-range) and without any form of interaction.  
Diagonalization can then be trivially achieved via Fourier transformation. Conserved charges can be directly
built from the operators representing occupation of the Fourier modes.  If we imagine for example a size sequence in which we double the
system size at each step and always use periodic boundary conditions, we can embed each such conserved charge into a sequence
of constant density character. We thus say that free hopping Hamiltonians are constant quantum integrable in the Fourier basis.
Looked at in the real-space basis, we can simply Fourier transform the conserved charges: these then become $O(N)$, and we can thus
say that free hopping Hamiltonians are linear quantum integrable in the sites basis.  

\subsubsection*{Fundamental models from quantum inverse scattering}
Going beyond free theories, the most straightforward class of models to discuss are those which fit within the scheme of the 
quantum inverse scattering method (see \cite{KorepinBOOK} and references therein), 
in other words which are treatable with the Algebraic Bethe Ansatz. This relies on the
existence of $R$-matrices solving the Yang-Baxter equation
\begin{equation}
R_{12} (\lambda,\mu) R_{13}(\lambda,\nu) R_{23}(\mu,\nu) = R_{23} (\mu,\nu) R_{13}(\lambda,\nu) R_{12}(\lambda,\mu).
\label{eq:YB}
\end{equation}
The existence of monodromy matrices $T (\lambda)$ obeying the intertwining relation
\begin{equation}
R_{12} (\lambda, \mu) T_1 (\lambda) T_2 (\mu) = T_2 (\mu) T_1 (\lambda) R_{12} (\lambda, \mu)
\label{eq:RTTequalsTTR}
\end{equation}
then allows to construct integrable models and their conserved charges by taking logarithmic
derivatives of the transfer matrix $\tau (\lambda) \equiv \mbox{Tr} ~T(\lambda)$:  the set of
operators
\begin{equation}
{\cal Q}_\alpha = c_\alpha \frac{d^\alpha}{d \lambda^\alpha} \ln \tau(\lambda) |_{\lambda = \xi}
\end{equation}
(where $\xi$ is some fixed evaluation parameter and $c_\alpha$ are arbitrary numerical constants) automatically constitutes an abelian set
by virtue of (\ref{eq:RTTequalsTTR}). It is reasonable to assume that the set of charges so obtained is maximal,
though this remains a conjecture. 
This logic allows to construct integrable quantum lattice models with local Hamiltonians by starting
from a monodromy matrix which is a product of local operators,
for example Heisenberg chains \cite{1979_Kulish_PLA_70} and more general models \cite{1983_Tarasov_TMP_57,1984_Tarasov_TMP_61}.
Interestingly, the search for a classification of all possible solutions to the Yang-Baxter equation
was initiated in works such as \cite{1981_Kulish_LMP_5,1982_Kulish_JMS_19} but remains open to this day. 

For the explicit case of the Heisenberg spin chain, 
\begin{equation}
H_{XXX} = J \sum_{j=1}^N {\bf S}_j \cdot {\bf S}_{j+1}
\label{eq:HXXX}
\end{equation}
(which has linear density character) the locality of the simplest conserved charges 
in the infinite system (here, locality means that the conserved charges can be expressed as
integrals of local densities in the continuum limit) was proven in \cite{1976_Luescher_NPB_117},
the structure being explicitly of the form
\begin{eqnarray}
{\cal Q}_n = \sum_{\{ j_1, ..., j_{n-1} \} } G^T_{n-1} (j_1, ..., j_{n-1}), \nonumber \\
G^T_n (j_1, ..., j_{n} ) = 0, \hspace{5mm} |j_1 - j_n | \geq n.
\end{eqnarray}
where the summation is over ordered subsets $\{j_1,\ldots,j_{n-1}\}$ of the chain and $G^T$ is a translationally 
invariant function.
The set of local charges (which, we remind the reader, {\it cannot} be the maximal set of 
conserved charges if $n/N \rightarrow 0$ where $N$ is the total number of sites) was 
shown in \cite{1979_Babbitt_JMAA_72} to be maximally abelian on the subspace of states 
where all but a finite number of spins point in the same direction. For finite sytems,
the detailed structure of the conserved charges was systematically studied in a series of papers by
Grabowski and Mathieu \cite{1994_Grabowski_MPLA_9,1995_Grabowski_JPA_28,1996_Grabowski_JPA_29},
and the explicit structure was found both for periodic and open boundary conditions. 
For example, for the $XXX$ chain, the simplest charge above the Hamiltonian is
\begin{equation}
{\cal Q}_3 = \sum_j {\bf S}_{j} \cdot ({\bf S}_{j+1} \times {\bf S}_{j+2}).
\label{eq:Q3XXX}
\end{equation}
A very interesting further paper by the same authors \cite{1995_Grabowski_JPA_28}
suggests to use the existence of a {\it single} higher local conserved charge as a litmus test
for integrability. The conjectures offered in this paper are that 1) a necessary condition for
integrability is the existence of a conserved charge made of terms coupling at most three sites,
and 2) a sufficient condition is the existence of a boost operator $B$ (in Baxter's logic \cite{BaxterBOOK}) generating
higher charges according to $\left[ B, {\cal Q}_n \right] = {\cal Q}_{n+1}$.  The authors recover many known
models using this logic, but while their ideas are very appealing and suggestive,
they unfortunately didn't produce new previously unknown cases. 

By looking at the structure of the conserved charges above, we can immediately see that 
each model obtained from a transfer matrix composed of products of quasi-local operators
has conserved charges having linear density character in the sites basis.
All of these are thus linear quantum integrable in the sites basis.
Note that a similar reasoning holds for inhomogeneous models, which can even be used to simulate disordered
systems \cite{1984_deVega_NPB_240}. The list of what we call linear quantum integrable models thus includes
many famous lattice models such as Heisenberg chains (also of higher spin), the $t-J$ and Hubbard models.
It also includes variants such as restricted Bose-Hubbard with up to two bosons per site, or
low-density limits of such lattice models yielding continuum counterparts like the 
Lieb-Liniger model \cite{2004_Amico_AP_314}.

\subsubsection*{Haldane-Shastry-type models}
An interesting generalization of the Heisenberg chain is the Haldane-Shastry long-range interacting model
\cite{1988_Haldane_PRL_60,1988_Shastry_PRL_60},
\begin{equation}
H_{HS} = \sum_{j_1, j_2 = 1}^N \frac{z_{j_1} z_{j_2}}{z_{j_1 j_2} z_{j_2 j_1}} {\bf S}_{j_1} \cdot {\bf S}_{j_2}
\end{equation}
in which $z_j \equiv e^{2\pi i j/N}$ and $z_{j_1 j_2} = z_{j_1} - z_{j_2}$. 
Since it involves
a double sum over sites of simple operator combinations, the Hamiltonian
of the Haldane-Shastry model is now an operator with quadratic density character. The conserved charges of the Haldane-Shastry
chain can be constructed explicitly \cite{1995_Talstra_JPA_28}. The first one above the Hamiltonian can be written
\begin{equation}
{\cal Q}_3 = i \sum_{j_1 j_2 j_3} \frac{z_{j_1} z_{j_2} z_{j_3}}{z_{j_1 j_2} z_{j_1 j_3} z_{j_2 j_3}} 
{\bf S}_{j_1} \cdot ({\bf S}_{j_2} \times {\bf S}_{j_3}).
\end{equation}
Since this charge involves three
space summations, it belongs to a size sequence with cubic density character. Each successive conserved charge
includes one additional summation over sites.  A specific conserved charge ${\cal Q}_n$ thus has density
character $O(N^n)$. Some care must thus be taken to characterize the class of integrability this model belongs to
according to our definition. The cardinality of 
the set of conserved charges becomes unbounded as $N \rightarrow \infty$, but a restricted set of charges up
to some ${\cal Q}_{n_{max}}$ can only be embedded in operator size sequences of density character $O(N^{n_{max}})$. 
To fulfill condition 3 of our definition, the Haldane-Shastry model must 
thus display a form of integrability which is distinct from the previous fundamental lattice models
(like {\it e.g.} the Heisenberg chain):  it cannot be a polynomial quantum integrable model. 
We can fit our definition of quantum integrability by choosing for example the density character 
$O(N^{\ln N}) \sim O(e^{\ln^2 N})$, which would make the Haldane-Shastry chain a quasi-polynomial
quantum integrable model.  

The Haldane-Shastry chain, which according to our definition thus displays a weaker form of 
integrability than {\it e.g.} the Heisenberg model,
is part of a more general family of long-range models 
of Calogero-Sutherland type \cite{1971_Calogero_JMP_12,1971_Sutherland_PRA_4,1972_Sutherland_PRA_5},
containing many generalizations \cite{1993_Polychronakos_PRL_70,1996_Inozemtsev_PS_53,1994_Frahm_JPA_27}.
Note that this class of models contains Heisenberg spin chains as a simple limit, and 
also more generally fits within the Yang-Baxter logic \cite{1993_Bernard_JPA_26},
but the transfer matrix now gives rise to the generators of a Yangian, a set of non-local charges which commute with all local conserved charges but not among themselves. Quasi-local conserved charges, including the one given above, 
can be obtained by expanding the so-called quantum determinant of the transfer matrix \cite{1995_Talstra_JPA_28}.

As mentioned before, the energy level statistics of Haldane-Shastry type models 
is neither Poissonian 
nor of Wigner type. We propose to understand this fact via the observation 
that this family of models fall in the class of quasi-polynomial quantum integrable, 
rather than linear or polynomial.

\subsubsection*{Richardson-Gaudin type models}
Yet another family of models which are interesting to discuss are the Richardson-like models
\cite{1963_Richardson_PL_3,1963_Richardson_PL_5,1964_Richardson_NP_52_1,GaudinBOOK,2002_von_Delft_PRB_66}. Similarly to the
Haldane-Shastry-type models, the Hamiltonian has quadratic density character. However, the conserved charges
are simpler \cite{GaudinBOOK}, and are of linear density character. We can thus put this family of models in the class of
linear quantum integrable in the sites basis. \\

Finally, in a generic model (which still fulfills {\bf QI:N}), it can be expected that
all charges will be of exponential density character. We would thus by convention say that
if a model is not at least sub-exponential integrable, it simply should be called non-integrable. \\

We could carry on mentioning other specific models. However, for the sake of compactness, and
given the fact that the examples we have given clearly illustrate our idea, we now turn to 
the more crucial question of the physical interpretation of the new definition.

\section{Physical meaning of the new definition}
\label{sec:Physics}
One of our requirements was that the classification scheme should be physically meaningful. 
We have already mentioned a few features traditionally associated with integrable models, such as factorized
scattering, nondiffraction, particular energy level statistics, {\it etc}. The main theme which we would
however like to understand better is how the classical notion of (non)ergodicity translates 
to the quantum case.

In this context, the appropriate starting point for discussion is probably Mazur's inequality \cite{1969_Mazur_P_43},
which has already proven useful in a number of cases, for example in \cite{1997_Zotos_PRB_55}.
Let us recall its statement \cite{1969_Mazur_P_43,1971_Suzuki_P_51,1973_Sirugue_P_65}.
Consider an operator $A$, and consider the canonical average $\langle \Delta A (0) \Delta A(t) \rangle$
where $\Delta A = A - \langle A \rangle$. 
If $\{ \bar{\cal Q}_\alpha \}$, $\alpha = 1, ..., N_{\bar{\cal Q}}$ is a set of constants of motion, the long-time average correlation obeys
\begin{equation}
\lim_{T \rightarrow \infty} \frac{1}{T} \int_0^T dt \langle \Delta A (0) \Delta A(t) \rangle \geq \sum_{\alpha = 1}^{N_{\bar{\cal Q}}}
\frac{| \langle (\Delta A) \bar{\cal Q}_\alpha \rangle|^2}{\langle \bar{\cal Q}_\alpha^2 \rangle}
\label{eq:Mazur}
\end{equation}
in which the charges are defined in such a way as to be `orthogonal' for the chosen form of averaging,
\begin{equation}
\langle \bar{\cal Q}_\alpha \bar{\cal Q}_\beta \rangle = \delta_{\alpha \beta} \langle \bar{\cal Q}_\alpha^2 \rangle
\label{eq:normalization}
\end{equation}
(note that we write the charges here as $\bar{\cal Q}_\alpha$ to distinguish them from the charges we have
used in Section \ref{sec:Definition}:  our charges do {\it not} depend on the averaging procedure chosen,
while the $\bar{\cal Q}_\alpha$ charges {\it do} by implementation of (\ref{eq:normalization})).
The inequality becomes an equality if the set of charges used is maximal.
The operator $A$ is called ergodic if 
\begin{equation}
\lim_{T \rightarrow \infty} \frac{1}{T} \int_0^T dt \langle \Delta A (0) \Delta A(t) \rangle = 0,
\label{eq:ergodicity}
\end{equation}
in other words if it is orthogonal to all the conserved charges. The physical picture is that all correlations 
have dissipated in Hilbert space, and no quasi-periodic behaviour can be identified.
Note the fact that in order to prove non-ergodicity, 
it is sufficient to find a {\it single} conserved charge having a finite overlap with
the operator of interest (more elaborately, an infinite set of vanishing contributions summing up to
a finite value), thereby making the right-hand side of (\ref{eq:Mazur}) nonvanishing.

How does our definition of integrability relate to Mazur's inequality? 
Let us make the following reasoning. 
We can first of all generalize Mazur's inequality to a generic energy-diagonal expectation value,
\begin{equation}
\langle ... \rangle \rightarrow 
\langle ... \rangle_{\bf f} \equiv \frac{1}{Z_{\bf f}} \sum_{\alpha} f_\alpha \langle \alpha | ... | \alpha \rangle,
\hspace{10mm} Z_{\bf f} \equiv \sum_{\alpha} f_\alpha
\end{equation}
in which $\alpha$ labels eigenstates and $f_{\alpha}$ are specified real parameters. The form of the
distribution is important: Mazur's inequality trivializes to an empty statement if the average is taken
over a single state, since the conserved charge eigenvalues then cancel in the numerator and denominator
in each term of the right-hand side. We thus focus our attention on distributions with at least some measure
of extensivity.

Let us consider physically meaningful operators $A$ which are represented by
simple matrices in the preferred basis. Such operators would thus be represented as
\begin{equation}
A = \sum_{i_1, ..., i_N} A_{i_1 ... i_{N_a}} {\bf e}^{i_1 ... i_{N_a}}
\end{equation}
with ${\mathbb N}_{\bf e} (A)$ (using the terminology of Section \ref{sec:Definition}) being 
for example such that $A$ is of polynomial density character (this would thus include all operators made of
finite products of on-site operators or the Fourier transform of such objects). 

Let us now consider computing the building blocks of Mazur's inequality, namely the expectation values
\begin{equation}
\langle (\Delta A) \bar{\cal Q}_\alpha \rangle_{\bf f}.
\label{eq:ratio}
\end{equation}
This is really a measure of the `overlap' of the operator $\Delta A$ with the conserved charges we choose to work
with. Following \cite{1971_Suzuki_P_51}, this can be translated at the operator level 
for an `orthogonal' (obeying (\ref{eq:normalization})) set of charges $\bar{\cal Q}_\alpha$. One can write
\begin{equation}
A = \sum_{\alpha} a_\alpha \bar{\cal Q}_\alpha + A'
\label{eq:AinQbasis}
\end{equation}
where $a_\alpha$ are c-numbers (implicitly depending on the averaging scheme chosen) 
and $A'$ is an operator which is off-diagonal in energy, 
and therefore doesn't contribute to Mazur's inequality.  The ratios in (\ref{eq:Mazur}) are then
simply related to the coefficients
\begin{equation}
a_\alpha = \frac{\langle (\Delta A) {\cal Q}_\alpha \rangle_{\bf f}}{\langle {\cal Q}_\alpha^2 \rangle_{\bf f}}.
\label{eq:aalpha}
\end{equation}

The next steps in the discussion depend on the integrability class our system falls into. 
Let us discuss the generic non-integrable case first. 
In this case, all the conserved charges ${\cal Q}_\alpha$ are expressed in the preferred basis
in terms of dense matrices with $O(\mbox{dim}({\cal H})^2)$ entries, and have thus exponential density character.
Pick any one of these charges, say ${\cal Q}_{\alpha_1}$, and use it as the basis
for constructing the set $\bar{\cal Q}_\alpha$ of orthogonal charges to be used.
The exponential density character of $\bar{\cal Q}_{\alpha_1}$ charge means that the coefficient $a_{\alpha_1}$ in 
(\ref{eq:AinQbasis}) will be exponentially small. Picking a second charge ${\cal Q}_{\alpha_2}$, using an
orthogonalization procedure to satisfy (\ref{eq:normalization}), one remains (except for utterly contrived cases) 
with an exponential density character
charge $\bar{\cal Q}_{\alpha_2}$ having vanishing overlap $a_{\alpha_2}$ with our operator $A$. This simply carries on,
and no charge can be found which overlaps with $A$, so Mazur's inequality indicates ergodicity.

Let us now turn to the case of an integrable model according to our definition (for example, 
a linear quantum integrable model, for which we have many charges ${\cal Q}_\alpha$ at our disposal which can be written as
single summations of local operator products). By their nature, these charges can be expected to have a finite overlap with 
operator $A$. Enforcing the orthogonality condition (\ref{eq:normalization}) does not change the density character
of the charges, so we obtain a set of orthogonal charges $\bar{\cal Q}_\alpha$ whose (summed) overlaps have the potential to give a finite value to the
right-hand side of Mazur's inequality. This must of course be checked in individual cases, but the door is clearly 
left open by the conserved charges' simple structure, related (in the sense of their simple density character) to that of operator $A$. 
The same conclusions would hold for other integrability classes,
for appropriate operators having a sufficient degree of similarity to the (simple) conserved charges.

Let us make some simple and obvious remarks. First of all, of course,
ergodicity only makes sense in the infinite system size limit. For a finite system, in parallel to the
naive definition {\bf QI:N} of quantum integrability, any quantum system will possess `reasonable' observables
failing to show ergodicity in Mazur's sense; these observables will however not be physically
meaningful for what we have referred to as non-integrable systems, since they must have exponential density
character.  One thus cannot say that a {\it system}
is ergodic or not, but rather only that a given observable $A$ is, for a given averaging in a given system. 
Our definition offers a way of understanding the origin of this ergodicity, or of its absence, by looking at
the detailed structure of the conserved charges, and their relationship with the observables considered within
the adopted averaging scheme. Our scheme of separating models into different integrability classes in fact leads
us to expect that ergodicity manifests itself differently in each of these classes. We will further characterize
this in subsequent publications.

Another context in which the role of quantum integrability is frequently discussed is the dynamics resulting 
from a non-equiblibrium initial state \cite{2010_Cazalilla_NJP_12,2010_Polkovnikov_10075331,2010_Rigol_10081930,2010_Relano_JSTAT_P07016}. 
A common way to bring a system out of equilibrium is by  means of a 
so-called quantum quench \cite{2007_Calabrese_JSTAT_P06008}. Consider a Hamiltonian $H_\lambda$ which depends on some parameter $\lambda$.  
The quantum quench now consists of preparing the system in an eigenstate of $H_\lambda$, followed by instantaneously 
changing the parameter $\lambda \rightarrow \lambda'$. If  $[H_\lambda,H_{\lambda'}]\neq 0$, one expects 
complicated non-equilibrium dynamics to ensue. To clarify this, let $|\phi\rangle$ be an eigenstate of $H_\lambda$ 
before the quench. We can now express $|\phi(t)\rangle$ for $t > 0$ 
as $e^{-it H_{\lambda'}} |\phi\rangle = \sum_{n} c_n |n\rangle e^{-i E_n t}$, where the summation is 
over all eigenstates of $H_{\lambda'}$ and the coefficients $c_n=\langle n | \phi\rangle$ represent the overlaps 
between the initial state $|\phi\rangle$ and the eigenstates $|n\rangle$ of $H_{\lambda'}$. The complicated form 
of $|\phi(t)\rangle$ makes it hard to make predictions at short time scales. 
It is common to simplify the problem by considering the long time average of an observable $A$, 
\begin{equation}
\lim_{T\rightarrow \infty} \frac{1}{T} \int_{0}^{T} \langle \phi(t) | A | \phi(t)\rangle = \sum_n |c_n|^2 \langle n | A | n\rangle
\end{equation}
which has the same energy-diagonal form as (\ref{eq:Mazur}).\footnote{As in Mazur's inequality, in 
general the summation also contains off-diagonal expectation values 
$\langle n| A |m \rangle$ as a result of degenerate energy levels. We neglect these here for simplicity.}
For most quenches, we can also make the assumption that the distribution $\{ |c_n|^2 \}$ is narrow in energy.
For a sufficiently large system one can try to approximate the distribution $\{ |c_n|^2 \}$  by a canonical ensemble
\begin{equation}
 \sum_n |c_n|^2 \langle n | A | n\rangle^2 = \frac{1}{Z}\mbox{Tr}\{ A e^{-\beta H}\}
\end{equation}
where the inverse temperature $\beta$ is determined from the initial condition
\begin{equation}
\langle \phi(0) | H | \phi(0)\rangle =\frac{1}{Z} \mbox{Tr} \left\{ H e^{-\beta H} \right\}
\end{equation}
and $Z$ is the usual partition function $Z= \mbox{Tr}\{e^{-\beta H}\}$. 
This approximation, while successful for some cases, fails in general for integrable systems,
this failure being typically attributed to the presence of additional conserved charges. 
It is thus natural and appealing to generalize the canonical ensemble by the so-called Generalized Gibbs ensemble \cite{2007_Rigol_PRL_98},
\begin{equation}
\langle A \rangle_{GGE} = \frac{1}{\mathcal{Z}} \mbox{Tr} \left\{ A \; e^{- \sum_{j} \beta_j {\cal Q}_j} \right\},
\end{equation}
the generalized partition function $\mathcal{Z}$ being now computed as $\mathcal{Z} = \mbox{Tr}{e^{-\sum_i \beta_i {\cal Q}_i}}$. 
Initial (quench-time) conditions fix the Lagrange multipliers $\beta_n$ via the self-consistency requirements
\begin{equation}
\langle \phi (0)| {\cal Q}_i |\phi (0) \rangle = \frac{1}{\mathcal{Z}} \mbox{Tr} \left\{ {\cal Q}_i \; e^{- \sum_{j} \beta_j {\cal Q}_j} \right\}.
\end{equation}
Using all available conserved charges as per the original formulation of this approach 
(in other words by adopting the {\bf QI:N} view of integrability), 
one might consider the eigenstate projectors ${\cal Q}_n = |n\rangle \langle n|$ as a maximal set of conserved charges to be exploited. 
While this would lead the an exact description, this is of limited use since 
in this case $e^{-\beta_n {\cal Q}_n} = |c_n|^2$ are simply the overlap coefficients mentioned above, this involving 
$O(dim(\mathcal{H}))$ parameters. 
What lacks in the description of the Generalized Gibbs ensemble is thus a prescription for which conserved charges should 
actually be included in the scheme.  
Since we assumed that the distribution $\{|c_n|^2\}$ is narrow in energy, one should include (besides the inevitable Hamiltonian) 
conserved charges that have a small variance in the distribution  $\{|c_n|^2\}$.
In contrast, if a conserved charge ${\cal Q}_n$ has large variance, its Lagrange multiplier $\beta_n$ would be negligible,
and including this charge in the scheme would not provide meaningful state selection. 
For an integrable system as defined in Section 4,  many conserved charges having a similar character as the Hamiltonian
are available (these having non-negligible Lagrange multipliers), and one cannot generally neglect their effect on the distribution. 
On the other hand in the non-integrable case most of the conserved charges will have an exponentially large variance 
for this particular distribution, and can thus be disregarded. Our proposed definition of quantum integrability
thus leads us to suggest using {\it restricted} generalized Gibbs ensembles, which we will return to in future work.

\section{Conclusions}
\label{sec:Conclusions}

One of the difficulties in discussing quantum integrability is of course that it can very
quickly feel at best confusing, and at worst rather pedantic. We feel that these Proceedings offered the
right kind of forum for this contribution, in view of the importance of an increased clarity of concepts 
for use in discussions about quantum many-body equilibration dynamics. 
Instead of attempting to define `quantum integrability', it might have been more reasonable and less confusion-inducing 
to simply introduce new terminology avoiding the keyword `integrability' altogether. We have tried to mitigate this risk
by suggesting to associate the terminology of algorithmic time complexity with any specific mention of quantum integrability.
If we say then that the Heisenberg chain is linear quantum integrable, it will be clear that the notions and definition 
of our paper are the ones that are referred to, and which consequences we associate to such a categorization.

One point which we should clarify is how we reconcile the current concepts with Integrable Field Theories (ITF) and
Conformal Field Theories. The point is that we have considered in detail how the morphology of the conserved charges of
a given model behaves as the system size is increased, {\it i.e.} as one moves towards a continuum/field theory limit.
If a model is classified as linear integrable in the sites basis within our scheme, it will automatically lead to
a field theory having an infinite number of conserved charges expressible as integrals of local densities by
choosing an $o(N)$ subset of charges in the finite lattice regularization. 

In the beginning, we listed our three main requirements: that a proper definition of quantum integrability should
be unambiguous, and should lead to different classes of models, these having different discernible types of physical behaviour. 
On this last point, while we have provided heuristic arguments, the specific and detailed consequences 
of our definition of quantum integrability have not yet been fully worked out to the level of general theorems. 
This first involves a rather immense amount of
work of looking at specific examples in detail, which we are pursuing at this time. This will teach us whether
our definition actually has interesting substance, which is perhaps not completely convincing at this point. 
While we think that our definition is not ambiguous, 
the level of mathematical rigour we have adopted is rather low in comparison to what is achievable in the field of
integrable models, but should be put in correspondence to the level of discussion on algorithmic time complexity.
We can thus be satisfied overall that our initial objectives feel as if they have been met.
Much work remains to be done, and we will extract more specific conclusions on particular classes of models and observables
in future publications.

\paragraph{Acknowledgements} The authors would like to thank S. Gravel, B. Nienhuis, B. Pozsgay 
and J. Stokman for useful discussions. Both authors
gratefully acknowledge support from the FOM foundation of the Netherlands.

\bibliographystyle{unsrt}
\bibliography{/Users/jscaux/WORK/BIBTEX_LIBRARY/JSCAUX_papers,/Users/jscaux/WORK/BIBTEX_LIBRARY/BOOKS,/Users/jscaux/WORK/BIBTEX_LIBRARY/OTHERS_TBP,/Users/jscaux/WORK/BIBTEX_LIBRARY/1926-1930,/Users/jscaux/WORK/BIBTEX_LIBRARY/1931-1935,/Users/jscaux/WORK/BIBTEX_LIBRARY/1936-1940,/Users/jscaux/WORK/BIBTEX_LIBRARY/1941-1945,/Users/jscaux/WORK/BIBTEX_LIBRARY/1946-1950,/Users/jscaux/WORK/BIBTEX_LIBRARY/1951-1955,/Users/jscaux/WORK/BIBTEX_LIBRARY/1956-1960,/Users/jscaux/WORK/BIBTEX_LIBRARY/1961-1965,/Users/jscaux/WORK/BIBTEX_LIBRARY/1966-1970,/Users/jscaux/WORK/BIBTEX_LIBRARY/1971-1975,/Users/jscaux/WORK/BIBTEX_LIBRARY/1976-1980,/Users/jscaux/WORK/BIBTEX_LIBRARY/1981-1985,/Users/jscaux/WORK/BIBTEX_LIBRARY/1986-1990,/Users/jscaux/WORK/BIBTEX_LIBRARY/1991-1995,/Users/jscaux/WORK/BIBTEX_LIBRARY/1996-2000,/Users/jscaux/WORK/BIBTEX_LIBRARY/2001-2005,/Users/jscaux/WORK/BIBTEX_LIBRARY/2006-2010}

\begin{thebibliography}{10}

\bibitem{ArnoldBOOK}
V.~I. Arnold.
\newblock {\em {Mathematical Methods of Classical Mechanics}}.
\newblock Springer, 1978.

\bibitem{1992_Weigert_PD_56}
S.~Weigert.
\newblock The problem of quantum integrability.
\newblock {\em Physica D: Nonlinear Phenomena}, 56(1):107 -- 119, 1992.

\bibitem{2004_Faddeev_HenriPoincare}
L.~D. Faddeev.
\newblock {What is complete integrability in quantum mechanics}.
\newblock {Proceedings of the Symposium Henri Poincar{\'e}, Brussels, 8-9
  October 2004}.

\bibitem{2009_Clemente_Gallardo_IJGMMP_6}
J.~Cl{\'e}mente-Gallardo and G.~Marmo.
\newblock {Towards a definition of quantum integrability}.
\newblock {\em Int. J. Geom. Meth. Mod. Phys.}, 6:129--172, 2009.

\bibitem{1954_Kolmogorov_DANSSSR_98}
A.~N. Kolmogorov.
\newblock {On Conservation of Conditionally Periodic Motions for a Small Change
  in Hamilton's Function.}
\newblock {\em Dokl. Akad. Nauk SSSR}, 98:527--530, 1954.

\bibitem{1963_Arnold_UMN_18}
V.~I. Arnold.
\newblock {Proof of a Theorem of A. N. Kolmogorov on the Preservation of
  Conditionally Periodic Motions under a Small Perturbation of the
  Hamiltonian.}
\newblock {\em Uspehi Mat. Nauk}, 18:13--40, 1963.

\bibitem{1962_Moser_NAWG_1}
J.~K. Moser.
\newblock {On Invariant Curves of Area-Preserving Mappings of an Annulus}.
\newblock {\em Nachr. Akad. Wiss. G{\"o}ttingen, II Math. Phys.}, 1:1--20,
  1962.

\bibitem{1984_Hietarinta_JMP_25}
J.~Hietarinta.
\newblock Classical versus quantum integrability.
\newblock {\em J. Math. Phys.}, 25(6):1833--1840, 1984.

\bibitem{2002_Gravel_JMP_43}
S.~Gravel and P.~Winternitz.
\newblock Superintegrability with third-order integrals in quantum and
  classical mechanics.
\newblock {\em J. Math. Phys.}, 43(12):5902--5912, 2002.

\bibitem{1931_vonNeumann_AM_32}
J.~von Neumann.
\newblock {{\"U}ber Funktionen Von Funktionaloperatoren}.
\newblock {\em Ann. Math.}, 32(2):191--226, 1931.

\bibitem{1995_Weigert_CSF_5}
S.~Weigert and G.~M{\"u}ller.
\newblock Quantum integrability and action operators in spin dynamics.
\newblock {\em Chaos, Solitons \& Fractals}, 5(8):1419 -- 1438, 1995.
\newblock Integrals, Entropy and Chaos.

\bibitem{1997_Braverman_TG_2}
A.~Braverman, P.~Etingof, and D.~Gaitsgory.
\newblock {Quantum integrable systems and differential Galois theory}.
\newblock {\em Transformation Groups}, 2:31--56, 1997.

\bibitem{1999_Horozov_LMP_49}
E.~Horozov and A.~Kasman.
\newblock {Darboux Transformations of Bispectral Quantum Integrable Systems}.
\newblock {\em Lett. Math. Phys.}, 49:131--143, 1999.

\bibitem{SutherlandBOOK}
B.~Sutherland.
\newblock {\em Beautiful Models}.
\newblock World Scientific, 2004.

\bibitem{1977_Berry_PRSLA_356}
M.~V. Berry and M.~Tabor.
\newblock {Level Clustering in the Regular Spectrum}.
\newblock {\em Proc. Roy. Soc. Lon. A}, 356:375, 1977.

\bibitem{1993_Montambaux_PRL_70}
G.~Montambaux, D.~Poilblanc, J.~Bellissard, and C.~Sire.
\newblock Quantum chaos in spin-fermion models.
\newblock {\em Phys. Rev. Lett.}, 70(4):497--500, 1993.

\bibitem{1993_Poilblanc_EPL_22}
D.~Poilblanc, T.~Ziman, J.~Bellissard, F.~Mila, and G.~Montambaux.
\newblock {Poisson vs. GOE Statistics in Integrable and Non-Integrable Quantum
  Hamiltonians}.
\newblock {\em EPL (Europhysics Letters)}, 22(7):537, 1993.

\bibitem{2004_Rabson_PRB_69}
D.~A. Rabson, B.~N. Narozhny, and A.~J. Millis.
\newblock {Crossover from Poisson to Wigner-Dyson level statistics in spin
  chains with integrability breaking}.
\newblock {\em Phys. Rev. B}, 69(5):054403, 2004.

\bibitem{2004_Relano_PRE_70}
A.~Rela\~no, J.~Dukelsky, J.~M.~G. G\'omez, and J.~Retamosa.
\newblock {Stringent numerical test of the Poisson distribution for finite
  quantum integrable Hamiltonians}.
\newblock {\em Phys. Rev. E}, 70(2):026208, 2004.

\bibitem{1004.4844}
G.~P. Brandino, R.~M. Konik, and G.~Mussardo.
\newblock {Energy Level Distribution of Perturbed Conformal Field Theories}.
\newblock arXiv:1004.4844.

\bibitem{2008_Stepanov_PRE_77}
V.~V. Stepanov, G.~M\"{u}ller, and J.~Stolze.
\newblock Quantum integrability and nonintegrability in the spin-boson model.
\newblock {\em Phys. Rev. E}, 77(6):066202, 2008.

\bibitem{2005_Finkel_PRB_72}
F.~Finkel and A.~Gonz\'alez-L\'opez.
\newblock {Global properties of the spectrum of the Haldane-Shastry spin
  chain}.
\newblock {\em Phys. Rev. B}, 72(17):174411, 2005.

\bibitem{2006_BasuMallick_NPB_757}
B.~Basu-Mallick and N.~Bondyopadhaya.
\newblock {Exact partition function of $SU(m|n)$ supersymmetric Haldane-Shastry
  spin chain}.
\newblock {\em Nucl. Phys. B}, 757(3):280 -- 302, 2006.

\bibitem{2008_Barba_EPL_83}
J.~C. Barba, F.~Finkel, A.~Gonz{\'a}lez-L{\'o}pez, and M.~A. Rodr{\'i}guez.
\newblock {The Berry-Tabor conjecture for spin chains of Haldane-Shastry type}.
\newblock {\em EPL (Europhysics Letters)}, 83(2):27005, 2008.

\bibitem{2008_Barba_PRB_77}
J.~C. Barba, F.~Finkel, A.~Gonz\'alez-L\'opez, and M.~A. Rodr\'\i{}guez.
\newblock {Polychronakos-Frahm spin chain of $BC_{N}$ type and the Berry-Tabor
  conjecture}.
\newblock {\em Phys. Rev. B}, 77(21):214422, 2008.

\bibitem{2009_Barba_NPB_806}
J.C. Barba, F.~Finkel, A.~Gonz{\'a}lez-L{\'o}pez, and M.A. Rodr{\'i}guez.
\newblock An exactly solvable supersymmetric spin chain of bcn type.
\newblock {\em Nucl. Phys. B}, 806(3):684 -- 714, 2009.

\bibitem{2010_Barba_NPB_839}
J.C. Barba, F.~Finkel, A.~Gonz{\'a}lez-L{\'o}pez, and M.A. Rodr{\'i}guez.
\newblock Inozemtsev's hyperbolic spin model and its related spin chain.
\newblock {\em Nucl. Phys. B}, 839(3):499 -- 525, 2010.

\bibitem{2010_Enciso_PRE_82}
A.~Enciso, F.~Finkel, and A.~Gonz{\'a}lez-L{\'o}pez.
\newblock {Level density of spin chains of Haldane-Shastry type}.
\newblock {\em Phys. Rev. E}, 82(5):051117, 2010.

\bibitem{1979_Babbitt_JMAA_72}
D.~Babbitt and L.~Thomas.
\newblock {Ground state representation of the infinite one-dimensional
  Heisenberg ferromagnet. IV. A completely integrable quantum system}.
\newblock {\em Jour. Math. Anal. Appl.}, 72(1):305 -- 328, 1979.

\bibitem{KorepinBOOK}
V.~E. Korepin, N.~M. Bogoliubov, and A.~G. Izergin.
\newblock {\em Quantum Inverse Scattering Method and Correlation Functions}.
\newblock Cambridge Univ. Press, 1993.

\bibitem{1979_Kulish_PLA_70}
P.~P. Kulish and E.~K. Sklyanin.
\newblock {Quantum inverse scattering method and the Heisenberg ferromagnet}.
\newblock {\em Phys. Lett. A}, 70(5-6):461 -- 463, 1979.

\bibitem{1983_Tarasov_TMP_57}
V.~O. Tarasov, L.~A. Takhtadzhyan, and L.~D. Faddeev.
\newblock {Local Hamiltonians for integrable quantum models on a lattice}.
\newblock {\em Theoretical and Mathematical Physics}, 57:1059--1073, 1983.

\bibitem{1984_Tarasov_TMP_61}
V.~O. Tarasov.
\newblock {Local Hamiltonians for integrable quantum models on a lattice. II}.
\newblock {\em Theor. Math. Phys.}, 61:1211--1215, 1984.

\bibitem{1981_Kulish_LMP_5}
P.~Kulish, N.~Reshetikhin, and E.~Sklyanin.
\newblock {Yang-Baxter equation and representation theory: I}.
\newblock {\em Lett. Math. Phys.}, 5:393--403, 1981.

\bibitem{1982_Kulish_JMS_19}
P.~P. Kulish and E.~K. Sklyanin.
\newblock {Solutions of the Yang-Baxter equation}.
\newblock {\em J. Math. Sci.}, 19:1596--1620, 1982.

\bibitem{1976_Luescher_NPB_117}
M.~L{\"u}scher.
\newblock {Dynamical charges in the quantized renormalized massive Thirring
  model}.
\newblock {\em Nucl. Phys. B}, 117(2):475 -- 492, 1976.

\bibitem{1994_Grabowski_MPLA_9}
M.~P. Grabowski and P.~Mathieu.
\newblock {Quantum integrals of motion for the Heisenberg spin chain}.
\newblock {\em Mod. Phys. Lett. A}, 9:2197, 1994.

\bibitem{1995_Grabowski_JPA_28}
M.~P. Grabowski and P.~Mathieu.
\newblock Integrability test for spin chains.
\newblock {\em J. Phys. A: Math. Gen.}, 28(17):4777, 1995.

\bibitem{1996_Grabowski_JPA_29}
M.~P. Grabowski and P.~Mathieu.
\newblock The structure of conserved charges in open spin chains.
\newblock {\em J. Phys. A: Math. Gen.}, 29(23):7635, 1996.

\bibitem{BaxterBOOK}
R.~Baxter.
\newblock {\em {Exactly Solved Models in Statistical Mechanics}}.
\newblock {Academic Press}, 1982.

\bibitem{1984_deVega_NPB_240}
H.~J. de~Vega.
\newblock Families of commuting transfer matrices and integrable models with
  disorder.
\newblock {\em Nucl. Phys. B}, 240(4):495 -- 513, 1984.

\bibitem{2004_Amico_AP_314}
L.~Amico and V.~Korepin.
\newblock {Universality of the one-dimensional Bose gas with delta
  interaction}.
\newblock {\em Ann. Phys.}, 314(2):496 -- 507, 2004.

\bibitem{1988_Haldane_PRL_60}
F.~D.~M. Haldane.
\newblock {Exact Jastrow-Gutzwiller resonating-valence-bond ground state of the
  spin-1/2 antiferromagnetic Heisenberg chain with 1/$r^{2}$ exchange}.
\newblock {\em Phys. Rev. Lett.}, 60(7):635--638, 1988.

\bibitem{1988_Shastry_PRL_60}
B.~S. Shastry.
\newblock {Exact solution of an S=1/2 Heisenberg antiferromagnetic chain with
  long-ranged interactions}.
\newblock {\em Phys. Rev. Lett.}, 60(7):639--642, 1988.

\bibitem{1995_Talstra_JPA_28}
J.~C. Talstra and F.~D.~M. Haldane.
\newblock {Integrals of motion of the Haldane-Shastry model}.
\newblock {\em J. Phys. A: Math. Gen.}, 28(8):2369, 1995.

\bibitem{1971_Calogero_JMP_12}
F.~Calogero.
\newblock {Solution of the One-Dimensional N-Body Problems with Quadratic
  and/or Inversely Quadratic Pair Potentials}.
\newblock {\em J. Math. Phys.}, 12(3):419--436, 1971.

\bibitem{1971_Sutherland_PRA_4}
B.~Sutherland.
\newblock {Exact Results for a Quantum Many-Body Problem in One Dimension}.
\newblock {\em Phys. Rev. A}, 4(5):2019--2021, 1971.

\bibitem{1972_Sutherland_PRA_5}
B.~Sutherland.
\newblock {Exact Results for a Quantum Many-Body Problem in One Dimension. II}.
\newblock {\em Phys. Rev. A}, 5(3):1372--1376, 1972.

\bibitem{1993_Polychronakos_PRL_70}
A.~P. Polychronakos.
\newblock {Lattice integrable systems of Haldane-Shastry type}.
\newblock {\em Phys. Rev. Lett.}, 70(15):2329--2331, 1993.

\bibitem{1996_Inozemtsev_PS_53}
V.~I. Inozemtsev.
\newblock {Exactly solvable model of interacting electrons confined by the
  Morse potential}.
\newblock {\em Phys. Scr.}, 53(5):516, 1996.

\bibitem{1994_Frahm_JPA_27}
H.~Frahm and V.~I. Inozemtsev.
\newblock {New family of solvable 1D Heisenberg models}.
\newblock {\em J. Phys. A: Math. Gen.}, 27(21):L801, 1994.

\bibitem{1993_Bernard_JPA_26}
D.~Bernard, M.~Gaudin, F.~D.~M. Haldane, and V.~Pasquier.
\newblock {Yang-Baxter equation in long-range interacting systems}.
\newblock {\em J. Phys. A: Math. Gen.}, 26(20):5219, 1993.

\bibitem{1963_Richardson_PL_3}
R.~W. Richardson.
\newblock A restricted class of exact eigenstates of the pairing-force
  hamiltonian.
\newblock {\em Phys. Lett.}, 3(6):277 -- 279, 1963.

\bibitem{1963_Richardson_PL_5}
R.W. Richardson.
\newblock Application to the exact theory of the pairing model to some even
  isotopes of lead.
\newblock {\em Phys. Lett.}, 5(1):82 -- 84, 1963.

\bibitem{1964_Richardson_NP_52_1}
R.~W. Richardson and N.~Sherman.
\newblock Exact eigenstates of the pairing-force hamiltonian.
\newblock {\em Nucl. Phys.}, 52:221 -- 238, 1964.

\bibitem{GaudinBOOK}
M.~Gaudin.
\newblock {\em La fonction d'onde de {B}ethe}.
\newblock Masson, Paris, 1983.

\bibitem{2002_von_Delft_PRB_66}
J.~von Delft and R.~Poghossian.
\newblock {Algebraic Bethe ansatz for a discrete-state BCS pairing model}.
\newblock {\em Phys. Rev. B}, 66(13):134502, 2002.

\bibitem{1969_Mazur_P_43}
P.~Mazur.
\newblock Non-ergodicity of phase functions in certain systems.
\newblock {\em Physica}, 43(4):533 -- 545, 1969.

\bibitem{1997_Zotos_PRB_55}
X.~Zotos, F.~Naef, and P.~Prelovsek.
\newblock Transport and conservation laws.
\newblock {\em Phys. Rev. B}, 55(17):11029--11032, 1997.

\bibitem{1971_Suzuki_P_51}
M.~Suzuki.
\newblock Ergodicity, constants of motion, and bounds for susceptibilities.
\newblock {\em Physica}, 51(2):277 -- 291, 1971.

\bibitem{1973_Sirugue_P_65}
M.~Sirugue and A.~Verbeure.
\newblock {On the lower bound of Mazur for autocorrelation functions.
  Ergodicity}.
\newblock {\em Physica}, 65(1):181 -- 186, 1973.

\bibitem{2010_Cazalilla_NJP_12}
M.~A. Cazalilla and M.~Rigol.
\newblock {Focus on Dynamics and Thermalization in Isolated Quantum Many-Body
  Systems}.
\newblock {\em New J. Phys.}, 12(5):055006, 2010.

\bibitem{2010_Polkovnikov_10075331}
A.~Polkovnikov, K.~Sengupta, A.~Silva, and M.~Vengalattore.
\newblock Nonequilibrium dynamics of closed interacting quantum systems.
\newblock {\em arXiv:1007.5331}, 2010.

\bibitem{2010_Rigol_10081930}
M.~Rigol.
\newblock Dynamics and thermalization in correlated one-dimensional lattice
  systems.
\newblock {\em arXiv:1008.1930}, 2010.

\bibitem{2010_Relano_JSTAT_P07016}
A~Rela{\~n}o.
\newblock Thermalization in an interacting spin system in the transition from
  integrability to chaos.
\newblock {\em J. Stat. Mech.: Th. Exp.}, 2010(07):P07016, 2010.

\bibitem{2007_Calabrese_JSTAT_P06008}
P.~Calabrese and J.~Cardy.
\newblock Quantum quenches in extended systems.
\newblock {\em J. Stat. Mech.: Th. Exp.}, 2007(06):P06008, 2007.

\bibitem{2007_Rigol_PRL_98}
M.~Rigol, V.~Dunjko, V.~Yurovsky, and M.~Olshanii.
\newblock {Relaxation in a Completely Integrable Many-Body Quantum System: An
  Ab Initio Study of the Dynamics of the Highly Excited States of 1D Lattice
  Hard-Core Bosons}.
\newblock {\em Phys. Rev. Lett.}, 98(5):050405, 2007.

\end{thebibliography}

\end{document}